\documentclass[%
reprint,
superscriptaddress,
amsmath,amssymb,
aps,
]{revtex4-1}

\newcommand {\Sec}[1] {Section~\ref{#1}}

\newcommand {\Fig}[1] {Figure~\ref{#1}}

\newcommand {\cc}{cm$^{-3}$}

\newcommand{\beq}{\begin{equation}}
\newcommand{\eeq}{\end{equation}}

\newcommand{\microsec}{$\upmu$s}

\newcommand{\natc}{N@C$_{60}$}

\newcommand{\csixty}{C$_{60}$}

\newcommand{\phosphorus}{$^{31}$P}

\newcommand{\ctwelve}{$^{12}$C}
\newcommand{\oneh}{$^{1}$H}
\newcommand{\cthirteen}{$^{13}$C}

\newcommand{\bismuth}{$^{209}$Bi}

\newcommand{\sitwonine}{$^{29}$Si}
\newcommand{\sitwoeight}{$^{28}$Si}

\newcommand{\beqa}{\begin{eqnarray}}
\newcommand{\eeqa}{\end{eqnarray}}

\newcommand{\JACS}{J. Am. Chem. Soc.}

\newcommand{\ttwo}{$T_2$}
\newcommand{\ttwostar}{$T_2^*$}

\newcommand{\tone}{$T_1$}

\newcommand{\ttwon}{$T_{\mathrm{2n}}$}

\usepackage{graphicx}
\usepackage{color}
\usepackage{dcolumn}
\usepackage{bm}
\usepackage{hyperref}
\usepackage[normalem]{ulem}

\usepackage{upgreek}

\begin{document}

\title{Storing quantum information in spins and high-sensitivity ESR} 

\author{John J. L. Morton}
\affiliation{London Centre for Nanotechnology, UCL, London WC1H 0AH, United Kingdom}
\affiliation{Dept. of Electronic and Electrical Engineering, UCL, London WC1E 7JE, United Kingdom}
\author{Patrice Bertet}
\affiliation{Quantronics group, SPEC, CEA, CNRS, Universit\'e Paris-Saclay, CEA Saclay 91191 Gif-sur-Yvette Cedex, France}

\date{\today} 

\begin{abstract}
Quantum information, encoded within the states of quantum systems, represents a novel and rich form of information which has inspired new types of computers and communications systems. Many diverse electron spin systems have been studied with a view to storing quantum information, including molecular radicals, point defects and impurities in inorganic systems, and quantum dots in semiconductor devices. In these systems, spin coherence times can exceed seconds, single spins can be addressed through electrical and optical methods, and new spin systems with advantageous properties continue to be identified. Spin ensembles strongly coupled to microwave resonators can, in principle, be used to store the coherent states of single microwave photons, enabling so-called microwave quantum memories. We discuss key requirements in realising such memories, including considerations for superconducting resonators whose frequency can be tuned onto resonance with the spins. Finally, progress towards microwave quantum memories and other developments in the field of superconducting quantum devices are being used to push the limits of sensitivity of inductively-detected electron spin resonance. The state-of-the-art currently stands at around 65 spins per $\sqrt{\rm Hz}$, with prospects to scale down to even fewer spins.
\end{abstract}

\maketitle

\newpage
%

\maketitle
	
	
\section{Introduction}
The storage of information using ferromagnetically-coupled spins in magnetic materials is commonplace through devices such as magnetic storage disks and MRAM, where as few as $\mathcal{O}(10^5)$ spins are used represent the logical `0' or `1' states of each bit~\cite{Nowak2016}. In contrast, the use of isolated, \emph{uncoupled} spins for information storage did not progress far beyond early studies in the 1950s~\cite{Anderson1955, Fernbach1955} --- these proposed a `spin echo serial storage memory' where weak NMR pulses were applied to a sample rich in nuclear spins to `write' data, which could then be recalled in arbitrary order using magnetic field pulses. Through such schemes, multiple bits could be stored as distinct collective excitations of an ensemble of uncoupled spins.
%
The storage capacity of such memories was determined by factors such as the thermal noise of the detecting apparatus, the effect of self-diffusion of the molecules, and relaxation times --- and although the `spin echo memory' may have promised some early advantages in latency against contemporary methods, we now know that these storage capacities were not able to become competitive with (ferro)magnetic information storage and its successors.

Through the concept of \emph{quantum} information~\cite{Deutsch1985,A.Nielsen2000} in the 1980s came the recognition that storing information in coherent quantum states offered the possibility of major, disruptive impacts in areas such as computing~\cite{Deutsch1985} and security~\cite{Bennett1985}. For example, a computer able to process quantum information, with a memory of only 50 quantum bits (`qubits'), would be able solve problems beyond the capabilities of today's most powerful supercomputers~\cite{Harrow2016}. In addition, quantum information, either stored or transmitted, can not be read by a third-party without detection, offering new methods for certification~\cite{Wiesner1983} and cryptography~\cite{Bennett1985}. Thus, storing information in isolated spins offers the possibility to exploit the \emph{coherence} of spins in fundamentally new ways and there is strong motivation to revisit the ideas of writing information into collective states of coherent spins. 

Inspired by such goals of using spins to represent qubits, extensive work has been performed over the past fifteen years on the measurement and extension of electron and nuclear spin coherence times (\ttwo) using a range of materials. Electron spin coherence times of seconds (at 6~K)~\cite{Wolfowicz2013} or milliseconds (at room temperature)~\cite{BarGill2013} have been measured, while nuclear spin coherence times in the solid-state can be as long as 40 minutes (room temperature)~\cite{Saeedi2013} hours to six hours (2~K)~\cite{Zhong2014}. Theoretical schemes have also been developed further for the storage and retrieval of quantum states within spin ensembles, using coupling to microwaves or light~\cite{Grodecka-Grad2012,Julsgaard2013}. In addition to serving as information storage elements for quantum information processors, such \emph{quantum memories} also play a key role in constructing so-called \emph{quantum repeaters}~\cite{Briegel1998,Sangouard2011}, which are required for to extend quantum communication over distances longer than a few hundred kilometers~\cite{Kimble2008}. 

In this Perspectives article, we review spins systems of interest to the storage of quantum information through quantum memories and survey their electron and nuclear spin coherence times. We will then consider methods for building quantum memories by coupling spins to high-Q superconducting resonators and examine the prospects for spin-based quantum information storage. Finally, we note that much of the technology around the development of quantum memories (such as high-Q superconducting resonators) can be combined with other developments arising from the field of quantum information processings (such as lossless quantum-limited microwave amplifiers) in order to yield significant advances in the sensitivity of electron spin resonance (ESR) detection. 

\begin{figure*}[t]
  \centering
  \includegraphics[width=140mm]{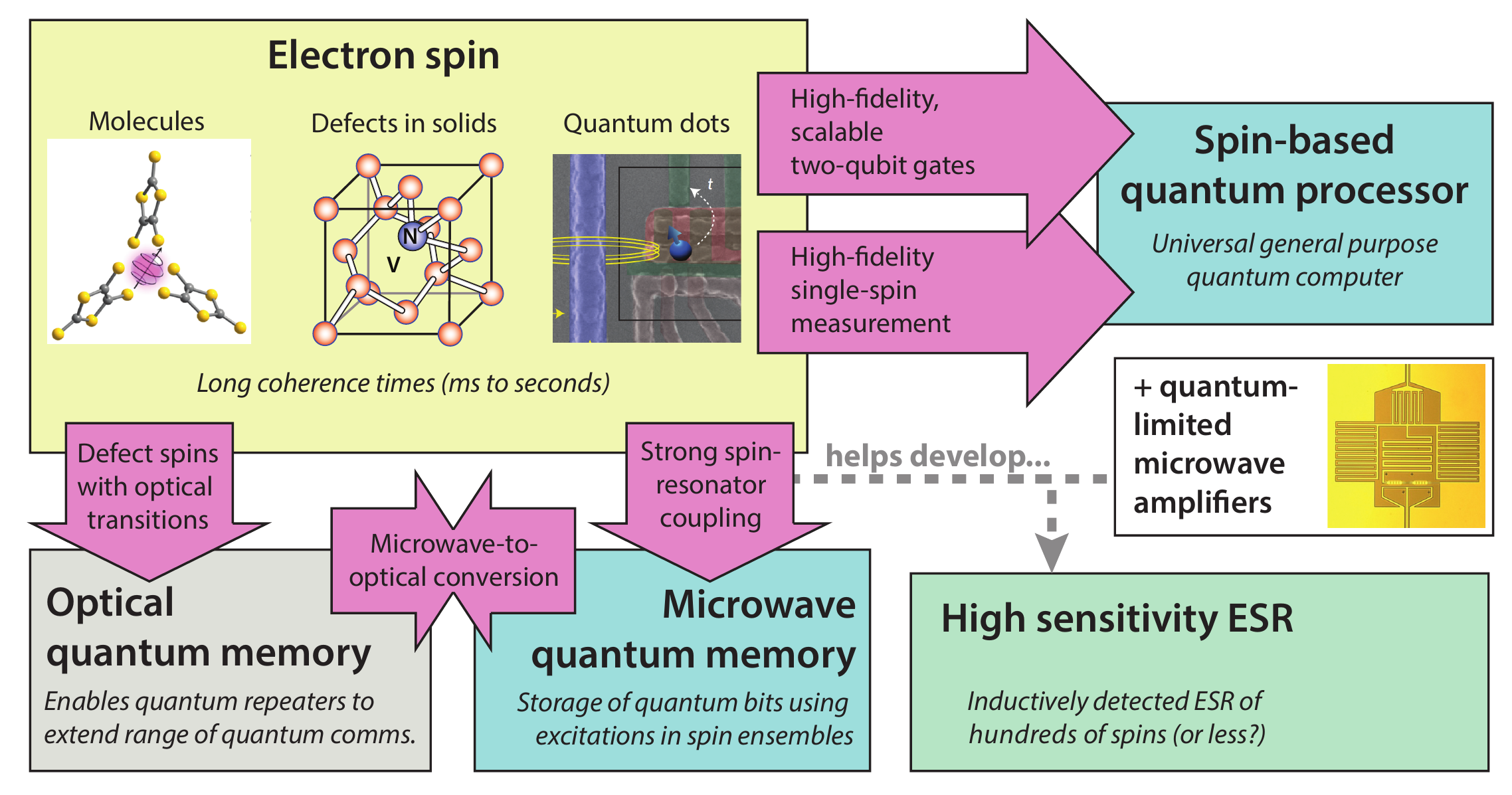}
  \caption{Summary of applications and requirements for quantum information storage with electron spins. Electron spin systems in a range of materials are being explored with applications in spin-based quantum information processing, microwave quantum memories, and optical quantum memories (though the latter can also be realised using the nuclear spin of non-Kramers ions). Here, we focus primarily on applications oriented towards microwave quantum memories, as well as opportunities in high-sensitivity ESR.
  }
  \label{fig:summary}
\end{figure*}

\section{Materials systems and spin coherence times}
\label{sec:systems}
A variety of electron spin systems are under investigation for the storage of quantum information --- these can broadly be divided into three categories: i)~those based on molecular systems which offer advantages such as tunability of spin properties through chemical design; ii)~those based on defect or impurity spins in inorganic solids where the crystalline environment and high purity can offer longer spin coherence times; and iii)~those based on the electron spins of artificial atoms, or quantum dots, which are electrostatically defined in semiconductors. To maximise electron spin coherence times, common goals across these different materials systems include minimising the local concentration of nuclear spins in order to suppress spectral diffusion~\cite{Balasubramanian2009,Tyryshkin2006}, as well as optimising the electron spin concentration in order to suppress the effects of electron-electron spin coupling (which leads to decoherence by instantaneous diffusion, and spectral diffusion~\cite{Tyryshkin2012})). 

Na\"ively, one might imagine that isolated spin-1/2 systems would be the most popular spin qubits, potentially offering longer coherence times to the presence of fewer decoherence mechanisms (e.g.~there is no crystal field or hyperfine tensor to modulate). In fact, as we shall see, the electron spins with the longest room-temperature coherence times happen to have $S\geq1$. Furthermore, by borrowing ideas from atomic clocks~\cite{Bollinger1985}, it is possible to identify spin transitions within some of these donors which have reduced sensitivity to common decoherence mechanisms (e.g.~arising from spin-spin interactions or magnetic field noise). These transitions are characterised by their value of $df/dB$ (the first order dependence of the transition frequency, $f$, with applied magnetic field $B$, equal to $g\mu_B/h$ for a free electron where $g$ is the electron g-factor, $\mu_B$ the Bohr magneton, and $h$ Planck's constant), which approaches zero at so-called `clock' or `ZEFOZ' (zero first-order Zeeman) transitions~\cite{Wolfowicz2013}. Such transitions are evidently absent for simple $S=1/2$ systems, but can result from the interplay of various spin Hamiltonian terms found in higher-spin systems or systems with strongly coupled electron and nuclear spins~\cite{Wolfowicz2013, Zhong2014, Zadrozny2017b,Dolde2011,Jamonneau2016}.

In addition to electron spins, nuclear spins have also been studied for quantum information storage --- indeed liquid state NMR was a powerful testbed for early ideas in quantum information. However, here we restrict ourselves to discussion of nuclear spins with the potential for strong coupling to other quantum degrees of freedom. For example, nuclear spins coupled to electron spins can offer memories based on longer nuclear spin coherence time~\cite{Morton2008} while retaining coupling to microwave resonators~\cite{Wu2010}, and nuclear spins coupled to optical transitions can be used for optical quantum memories and repeaters~\cite{Zhong2014}.

\begin{figure*}
  \centering
  \includegraphics[width=120mm]{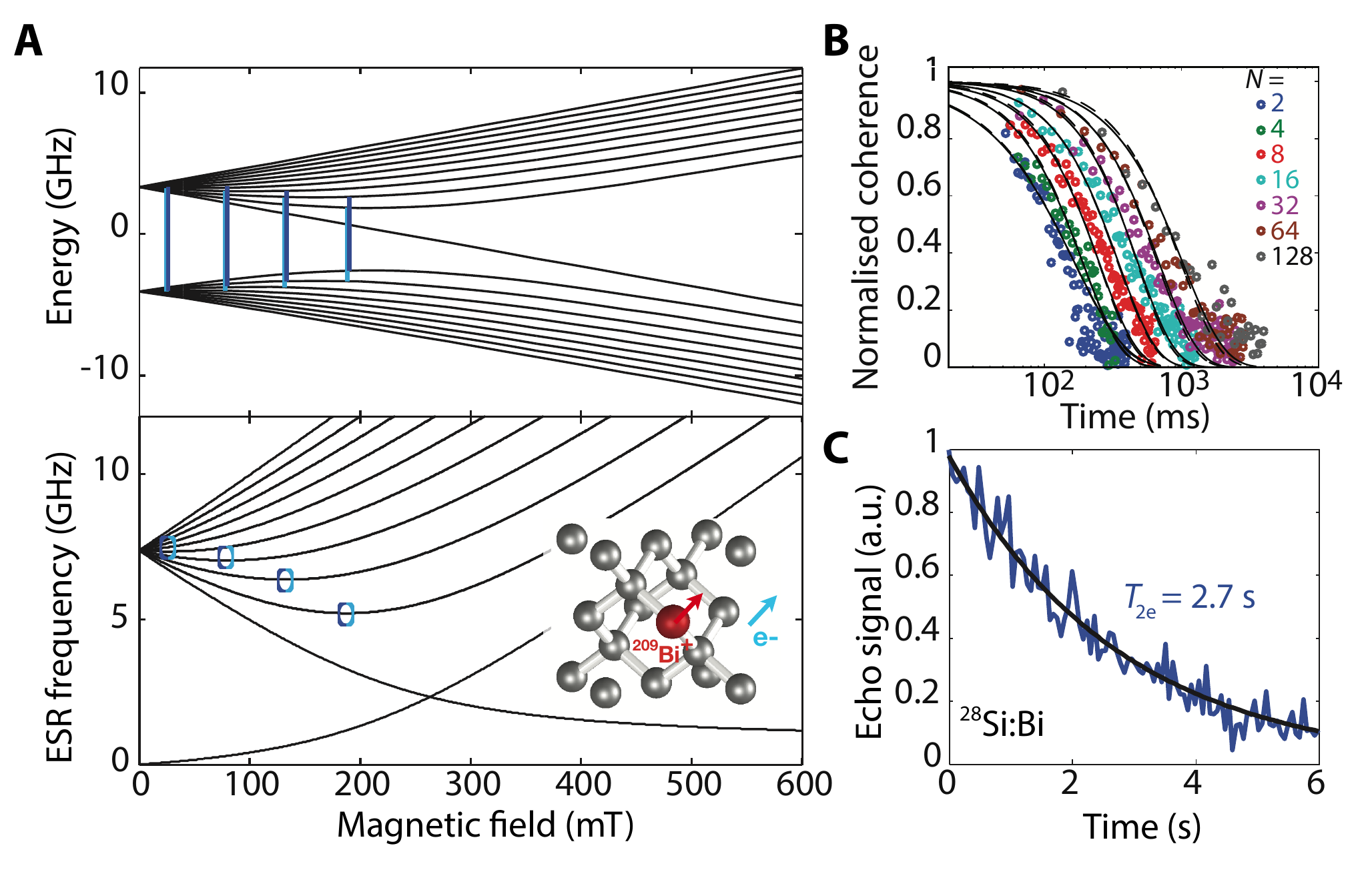}
  \caption{(A) Spin energy level spectrum and ESR transition frequencies of Bi donors in Si, as a function of magnetic field, highlighting four ESR clock transitions. (B) Electron spin coherence time measurements made using (top) natural silicon ([Bi]~$=3\times10^{15}$~\cc) under various degrees of CPMG dynamical decoupling and (bottom) isotoptically engineered \sitwoeight\ ([Bi]~$=4\times10^{14}$~\cc). 
  Adapted with permission from Ref.~\cite{Wolfowicz2015a}, \textcopyright~American Physical Society, and Ref.~\cite{Wolfowicz2013}, \textcopyright~(2013) Macmillan Publishers Ltd: Nature Nanotechnology.
   }
  \label{fig:bismuth}
\end{figure*}

\subsection{Molecular electron spins}
In molecular systems, the choice of solvent and ligand environment plays a critical role in determining the spin coherence time. Deuterating both solvents and molecular environments offers substantial improvements in electron spin coherence time, as has been exploited in double electron-electron resonance (DEER) experiments~\cite{Ward2010}, however, the longest coherence times are measured in environments that are essentially free of nuclear spins, achieved using solvents such as CS$_2$. For example, the endohedral fullerene molecule \natc~\cite{Harneit2002,Mehring2004}, in which a nitrogen atom is trapped within a C$_{60}$ fullerene cage, offers an $S=3/2$ electron spin with a coherence time of 80~\microsec\ in CS$_2$ at room temperature, rising to 240~\microsec\ at 170~K~\cite{Morton2006a}, just above the freezing point of the solvent. CS$_2$ crystallizes upon freezing, leading to the segregation of the \natc\ into regions of high concentration. The $^{15}$N nuclear spin of the \natc\ can be used to coherently store one of the electron spin coherences within the $S=3/2$ manifold, offering a \ttwon~$> 100$~ms at 10~K in a \csixty~matrix~\cite{Brown2011}.

ESR clock transitions have been identified and studied in molecular magnets as a way to suppress decoherence from nuclear spins or spin-spin coupling in samples with large concentrations~\cite{Zadrozny2017b, Shiddiq2016}, though coherence times in such cases remain rather low (between 8 and 14~\microsec). More promising, in terms of \ttwo, has been the synthesis of a molecular magnet with ligands that are nuclear-spin-free and offer good solubility in CS$_2$~\cite{Zadrozny2015}. In this way, segregation of the molecule upon solvent freezing is avoided, enabling the study of dilute molecules in a nuclear-spin-free matrix down to low temperatures, and the vanadium(IV) complex showed an electron spin coherence time of 700~\microsec\ at 10~K. Despite such impressive progress, the unique advantages of molecular systems remain to be fully demonstrated --- as we shall see, their coherence times fall well short of what can be achieved using certain inorganic materials, and the exciting potential of synthetic chemistry to achieve complex structures of coupled spins~\cite{Zadrozny2017} must be tempered with challenges such as maintaining chemical purity and molecular orientation across a spin ensemble, the punishing restrictions of using nuclear-spin-free ligands, and the absence of a straightforward route to single spin measurement.

\subsection{Electron spins in inorganic solids}	
\subsubsection{Donor spins in silicon}

Thanks to the key role which it plays in the information technology industry and the resulting work on perfecting high-purity crystal growth, silicon has become one of the purest substances mankind has produced. Impurity concentrations below 10$^{13}$~\cc\ (or 0.2 parts per billion) are routine in silicon, and isotopic enrichment of the majority \sitwoeight\ isotope ($I=0$) has been performed with a remaining concentration of the \sitwonine\ isotope $(I=1/2)$ down to 50~ppm~\cite{Andreas2011}. When silicon is doped with Group V elements, or `donors', such as P, As, Sb or Bi, and cooled below about 50~K, electrons become bound to the donor atom, with an isotropic hyperfine coupling to the donor nucleus. The electron spins of such donors have coherences times which are limited to several hundreds of microseconds in natural abundance silicon (5$\%$ \sitwonine)~\cite{Tyryshkin2006,George2010}, but can reach tens of milliseconds in \sitwoeight\ with $[P] = 10^{14}$~\cc, limited only by instantaneous diffusion caused by the finite concentration of donor electron spins~\cite{Tyryshkin2012}.

The Bi donor, where the $S=1/2$ electron spin is coupled to the ($I=9/2$) \bismuth\ nuclear spin, was used to perform the first demonstration of clock transitions in ESR~\cite{Wolfowicz2013}, achieving electron spin coherence times of up to 2.7~s at 6~K in enriched \sitwoeight, measured using a Hahn echo. At such clock transitions, \ttwo\ increases to 100~ms in natural abundance silicon, while dynamical decoupling techniques such as CPMG~\cite{Meiboom1958} can be used to extend this to approach a second~\cite{Wolfowicz2015a} (see \Fig{fig:bismuth}). From the perspective of quantum memories based on spin ensembles, the most significant advantage of using ESR clock transitions in Bi is the ability to increase the spin concentration by up to two orders of magnitude (and thus increase the strength of the spin-resonator coupling by a factor of $\mathcal{O}(10)$, as discussed in \Sec{sec:qmem}), whilst maintaining the same spin coherence lifetime.

Additional features which make donor spins in silicon attractive for quantum information storage include the ability to place them with near-atomic precision using scanning tunnelling microscopy (STM)-based hydrogen lithography~\cite{Watson2017}, while the electron and nuclear spin states of single donor spins can be read out with high fidelity by incorporating the donor into a silicon nanoelectronic device~\cite{Pla2012,Pla2013}. The coherent state of the donor electron spin can be stored in the nuclear spin~\cite{Morton2008}, giving access to coherence times of up to 180~s at a nuclear spin clock transition of the neutral \phosphorus\ donor~\cite{Steger2012}. 

\subsubsection{Optical spin defects in diamond and SiC}
Since it was first studied in the context of quantum information processing~\cite{Jelezko2004a}, the nitrogen-vacancy (NV) centre in diamond has stimulated a dramatic growth in research across a range of applications, now spanning magnetic-field, electric-field and temperature sensing and imaging down to the cellular and molecular level. One key feature the NV centre offers is the ability to measure the spin state of a single defect through changes in the optical photoluminescence~\cite{Jelezko2004a}, even at room temperature. Furthermore, the large Debye temperature of diamond, coupled with the mainly nuclear-spin-free carbon environment leads to electron spin coherence times of several hundred microseconds at room temperature, extendable to a few milliseconds using CPMG~\cite{BarGill2013}, or by using \ctwelve-enriched diamond~\cite{Balasubramanian2009}. While this significant room-temperature coherence opens up a range of sensing applications, the use of NV-centres in diamond for quantum information is likely to require cryogenic temperatures: to be used as a single-spin or ensemble microwave quantum memory, the temperature must be low enough to suppress blackbody radiation at the microwave frequency, $\omega$ (such that $T \ll \hbar \omega/k_B$, where $k_B$ is Boltzmann's constant). Furthermore, high-fidelity single-shot readout and spin-photon entanglement exploit spin-selective optical transitions which can only be resolved at low temperatures~\cite{Robledo2011}. The nitrogen nuclear spin of the NV centre can be used to store the state of electron spin, as can \cthirteen\ in the environment around the defect which can be  manipulated by periodically flipping (e.g.~by CPMG) the NV centre electron spin at the NMR frequency~\cite{Taminiau2014}.

Other centres in diamond, with potential advantages over the NV, are the subject of very active investigation. The negatively charged silicon-vacancy SiV$^{-1}$ ($S=1/2$, \ttwo~$\sim 100$~ns at 3.6~K)~\cite{Pingault2017,Rogers2014} and germanium-vacancy GeV$^{-1}$ (\ttwostar~$\sim 20$~ns at 2.2~K)~\cite{Siyushev2016, Bhaskar2017} centres have strong coherent optical transitions, but rather short coherence times at liquid helium temperatures due to their orbital degeneracy. 
However, recent studies of SiV$^{-1}$ in diamond have shown considerable increases in coherence properties when cooled down to 10~mK, with \ttwostar~$\sim10~\upmu$s and \ttwo\ up to a several hundred microseconds~\cite{Becker2017,Sukachev2017}.
The neutral SiV$^0$ centre ($S=1$), in contrast, has \ttwo~$\sim1$~ms at 20~K, while still retaining most (90\%) of its optical emission in the zero phonon line, a combination which makes it very promising~\cite{Rose2017}. Finally, silicon carbide (SiC) is another host offering a largely nuclear-spin-free environment, and defect centres such as the neutral ($kk$)-divacancy in $4H$-SiC ($S=1$, \ttwo~$\sim 1$~ms at 20~K)~\cite{Seo2016} are being investigated. 

\begin{figure}
  \centering
  \includegraphics[width=85mm]{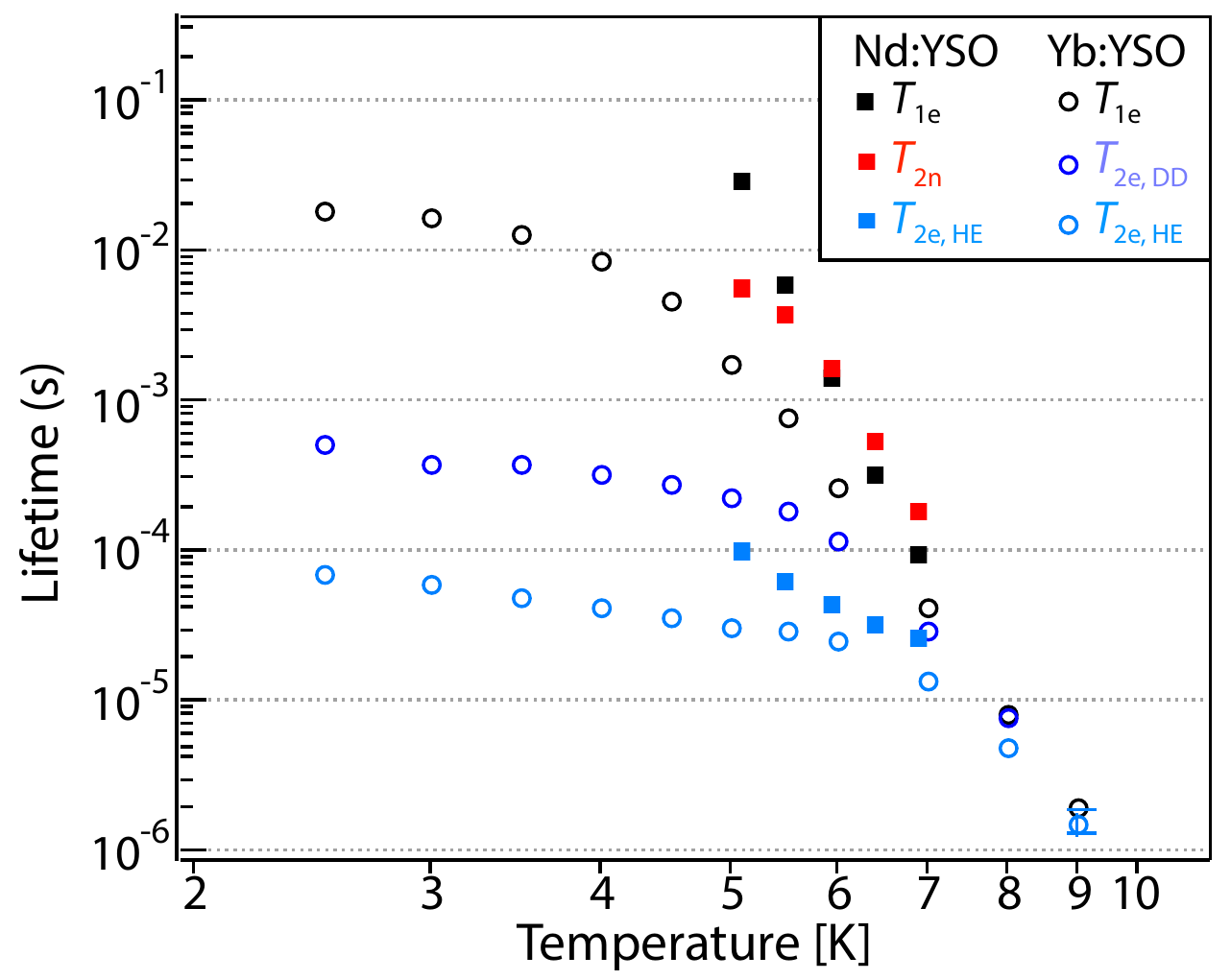}
  \caption{Spin coherence and relaxation time measurements of Nd and Yb ions in YSO, illustrating features such as the ability to exploit the longer coherence time of the nuclear spin (\ttwon), and the ability to extend the electron spin coherence time using dynamical decoupling (such as XY16) to suppress effects of spectral diffusion from non-resonant electron spins. Measurements on Yb were performed using $^{171}$Yb in site I at a magnetic field of 1020.8~mT, addressing the $m_I=-1/2$ ESR transition~\cite{Lim2017}. Measurements on Nd were performed using $^{145}$Nd at a magnetic field of 561.5~mT addressing the $m_I=+7/2$ ESR transition. 
Adapted with permission from Ref.~\cite{Wolfowicz2015}, \textcopyright~American Physical Society. }
  \label{fig:RE}
\end{figure}

\begin{figure*}
  \centering
  \includegraphics[width=140mm]{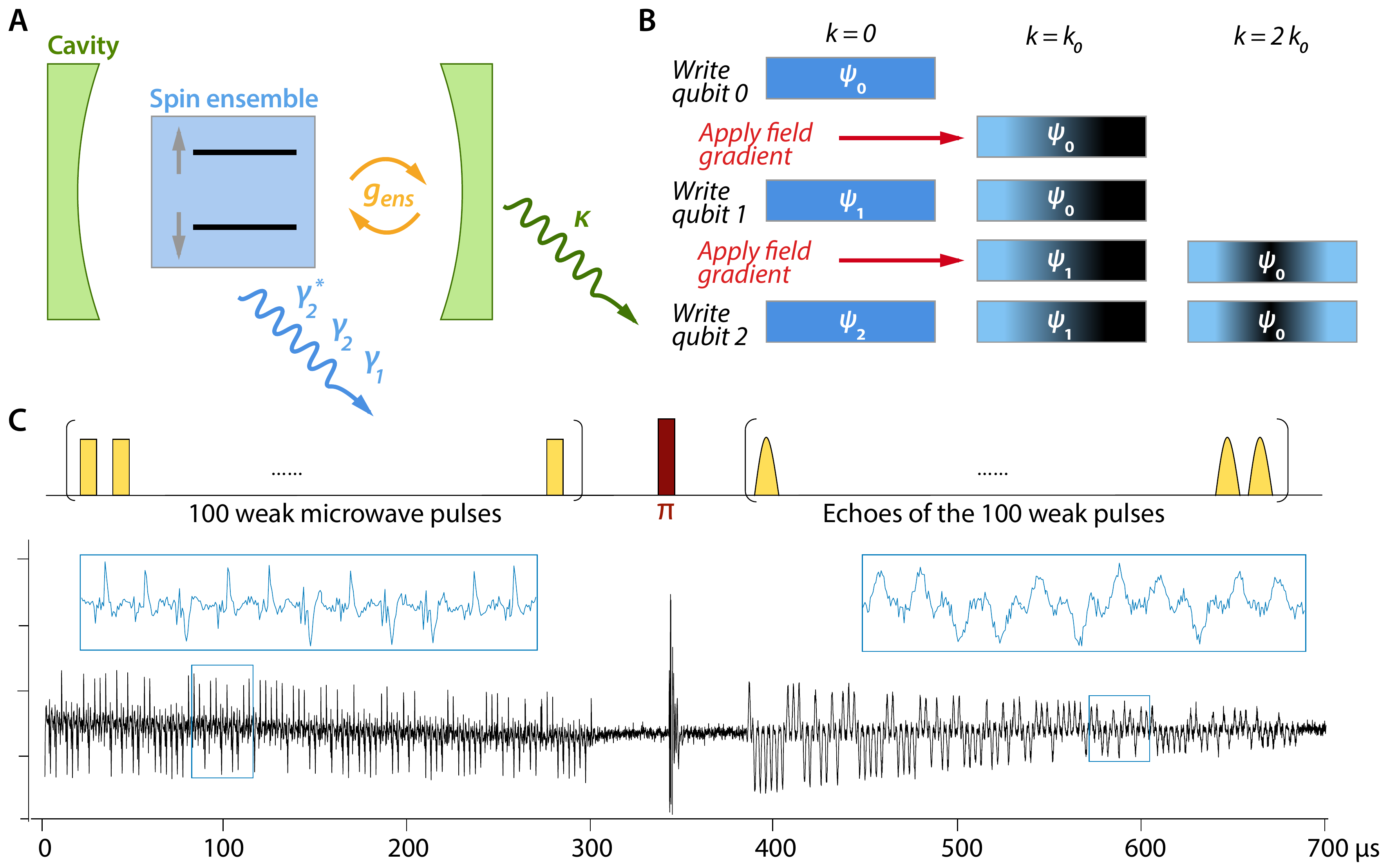}
  \caption{(A) Summary of key terms in cavity quantum electrodynamics with spin ensembles. $g$ is the spin-cavity coupling strength (which scales with $\sqrt(N)$ for $N$ spins) and $\kappa$ is the cavity damping rate (which can be expressed as $\omega_c/Q$, where $Q$ is the cavity Q-factor and $\omega_c$ the cavity frequency). The terms $\gamma_2^*$, $\gamma_2$ and $\gamma_1$ refer to the various decay rates for the spin ensemble: respectively, the inhomogeneous ensemble dephasing rate ($=1/T_2^*$), the Hahn-echo ensemble decoherence rate ($=1/T_2$) and the spin-lattice relaxation rate ($=1/T_1$). (B) Protocol for multi-mode microwave quantum memory using a spin ensemble. Single microwave photon states can be stored as collective excitations in the spin ensemble ($k=0$ mode), by allowing the spins and cavity to resonantly couple for some time ($\pi/g$). By applying a magnetic field gradient, or relying on intrinsic inhomogenously broadening, this collective excitation is transferred to some other mode, $k_0$, allowing a second quantum state to be written to the $k=0$ mode of the same ensemble. This process can be repeated allowing $\mathcal{O}(\sqrt{N})$ states to be written. (C) An illustration of the multi-mode memory concept, performed using weak microwave pulses (containing large numbers of microwave photons) and P-donors in silicon.
Panel (C) reprinted with permission from Ref~\cite{Wu2010}, \textcopyright~(2010) American Physical Society.
 }
  \label{fig:qmemory}
\end{figure*}

\subsubsection{RE ions in optical crystals}
The defect centres in diamond described above have optical transition wavelengths in the range of 600~nm (GeV$0$) to 950~nm (SiV$^0$), and, with the exception of the recently explored SiV$^0$, those with the best optical properties have shown relatively poor spin coherence times. Another class of spin defect of importance to quantum information storage is that of rare-earth (RE) dopants in crystal hosts such as YSO (yttrium orthosilicate) and YLiF$_4$. Such defects offer optical transition wavelengths up to 1550~nm (a technologically significant value to its use in optical fibre communication). RE dopant spins can be divided into two classes: those of Kramers ions (e.g.~Er, Nd, Yb) which have ESR transitions and can couple to microwave fields, and those of non-Kramers (e.g.~Eu, Pr) ions where the nuclear spin can be used alongside the optical transition to create an optical quantum memory~\cite{Hedges2010}. The non-Kramers ions offer the longest spin coherence times (\ttwon\ can be up to six hours using a nuclear spin clock transition~\cite{Zhong2014}) due to the absence of decoherence via an electron spin. However, use of Kramers ions is necessary to operate at optical wavelengths of 1550~nm (i.e.\ using Er), or as a route to develop quantum coherent microwave-to-optical conversion based on microwave and optical quantum memories operating in tandem~\cite{OBrien2014,Chen2016}\footnote{Coherent conversion of single microwave photons to optical photons would enable important applications such as the deterministic entanglement of optical photons, or building room temperature optical links between quantum processors operating in separate cryostats. Methods to achieve such conversion following a number of different designs are currently being pursued.}. Recently, nuclear spin coherence times over one second have been measured in Er-doped YSO, by applying high fields (7~T) and low temperatures (1.4~K) to suppress electron spin dynamics~\cite{Rancic2016}, in contrast to ENDOR measurements on Nd at lower field which showed \ttwon~up to only 9~ms~\cite{Wolfowicz2015}.
Electron spin coherence times of RE spins have been measured for Nd, Er, and Yb to be up to 100~\microsec\ (see \Fig{fig:RE})~\cite{Wolfowicz2015, Bertaina2007, Lim2017}, and should be compared with the electron g-factor (which can vary from 0.7 to 14) when assessing suitability for strong coupling to microwave resonators (see \Sec{sec:qmem}).


\subsection{Electron spins of quantum dots}
In addition to the `natural' spins of molecules, atoms and defects described above, electron spins of `artificial' quantum dot (QD) systems are attractive for hosting quantum information, particularly from the point of view of engineering scalable spin-spin interactions to build up quantum information processors. Such QDs have been studied in ensembles by conventional ESR, showing \ttwo~= 380~\microsec\ for QDs formed within Si/SiGe heterostructures~\cite{Jock2015}. There is also growing work at the single spin level based on electrical spin readout, with coherence times measured up to 1~ms (by Hahn echo) for QD spins formed within isotopically purified \sitwoeight\ MOS nanodevices~\cite{Veldhorst2014}. A promising future direction is developing electrical driving of spin resonance (EDSR) in such structures, as microwave electric fields are more readily confined at the nano-scale than the microwave magnetic fields used for conventional ESR. One method to achieve this is to use a double QD and magnetic field gradient, to create an effective spin-orbit coupling to allow EDSR --- such control has been demonstrated in Si/SiGe QDs with a cobalt micromagnet used to provide a field gradient, and coherence times up to 40~\microsec\ have been measured~\cite{Kawakami2014}. Finally, coupled systems of an impurity spin and QD spin have been studied and offer a hybrid implementation of a singlet-triplet spin qubit~\cite{Urdampilleta2015}.
	

\begin{figure*}
  \centering
  \includegraphics[width=150mm]{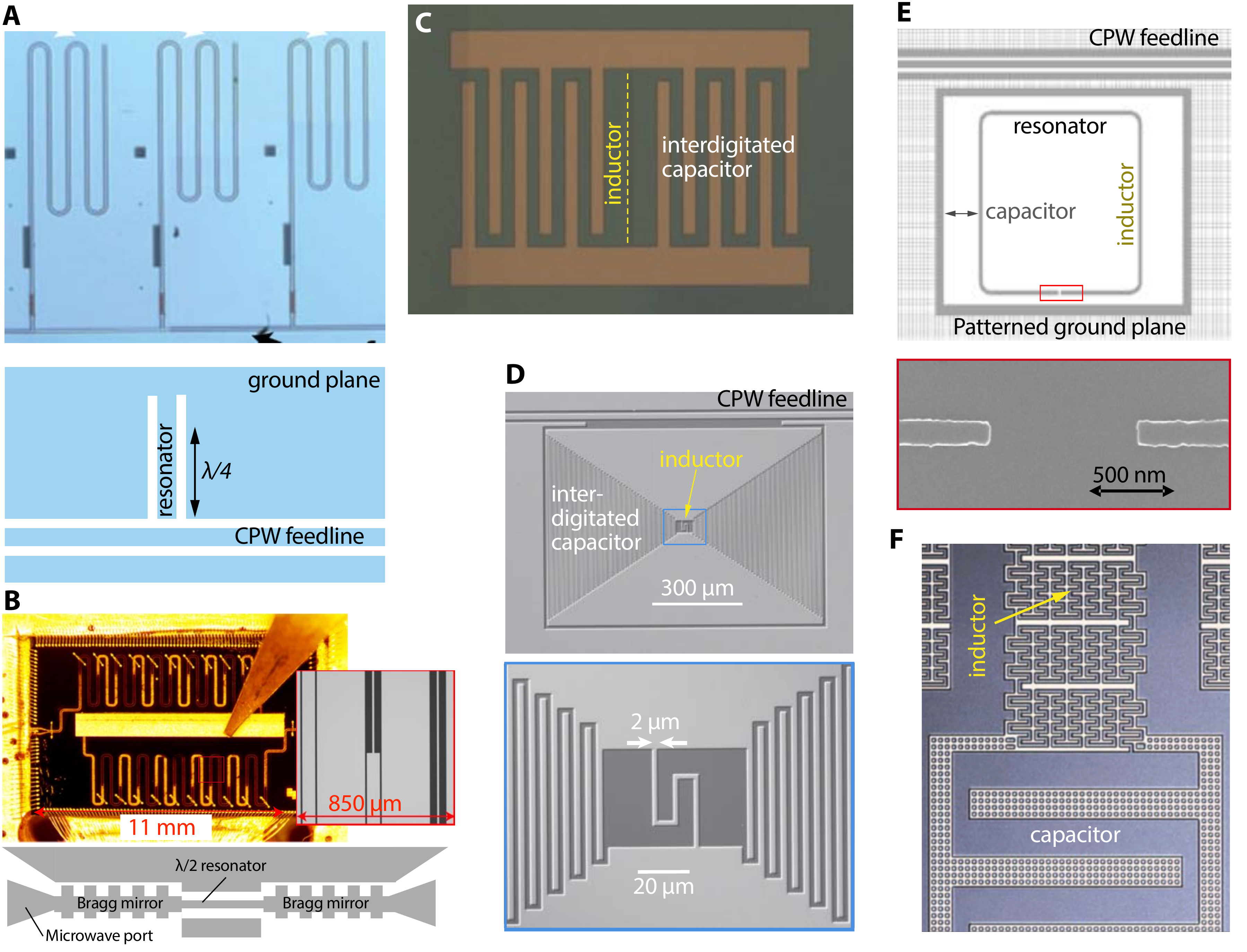}
  \caption{An overview of types of superconducting thin-film resonators designed for coupling to spin ensembles. 
  (A) Co-planar waveguide (CPW) resonators of length $\lambda/4$ (deposited on Al$_2$O$_3$), coupled to a CPW feedline, with optical micrograph (top) and schematic (bottom), adapted with permission from Ref.~\cite{Schuster2010}, \textcopyright~American Physical Society (APS). 
  (B) CPW resonator of length $\lambda/2$ defined by Bragg mirrors formed by modulating the width (and thus, impedance) of the transmission line, all patterned from a Nb film on Si. The panel shows an optical micrograph, with detailed inset (top) and schematic (bottom), and is
  adapted by permission from Macmillan Publishers Ltd: Nature Nanotechnology, Ref.~\cite{Sigillito2017}, \textcopyright~(2017).
  (C) Lumped element LC resonator made of Al on Si, coupled to microwave antennae via a copper box cavity (not shown)~\cite{Bienfait2016}.
  %
  (D) Low-impedance lumped element LC resonator made of Nb on Si, adapted with permission from Ref.~\cite{Eichler2017}, \textcopyright~APS. 
  (E) High-kinetic inductance resonator made of patterned NbTiN (where the dominant capacitance is that between the patterned loop and the ground plane) capable of maintaining $Q>2\times10^5$ in a 6~T in-plane magnetic field. Adapted with permission from Ref.~\cite{Samkharadze2016}, \textcopyright~APS.
  (F) Fractal resonator designed in order to tolerate high out-of-plane magnetic fields, demonstrating $Q> 25 000$ with a 20~mT field perpendicular to the film.
Reprinted from Ref.~\cite{Graaf2012} with the permission of AIP Publishing. 
  }
  \label{fig:resonators}
\end{figure*}

\section{Quantum memories}
\label{sec:qmem}
While the most promising schemes for spin-based quantum information \emph{processing} require qubits represented by single spins~\cite{Pla2012,Veldhorst2014,Kawakami2014}, many attractive realisations for quantum information \emph{storage} make use of ensembles: As with the spin echo serial storage memory proposals of the 1950s, the qubit is represented as a collective excitation of the ensemble. 
Key ideas around quantum information storage using ensembles were extensively developed and explored in the context of \emph{optical} quantum memories in which the states of single optical photons are stored in, and subsequently retrieved from, an ensemble of optical emitters, often making use of their internal hyperfine states~\cite{Grodecka-Grad2012}. Such optical memories are a pre-requisite for so-called quantum repeaters~\cite{Briegel1998,Sangouard2011,Muralidharan2016} which are used to distribute entangled states of light over long distances and underpin a quantum-secure communication network~\cite{Kimble2008}.
Here, we focus on \emph{microwave} quantum memories~\cite{Rabl2006,Julsgaard2013,Grezes2016}, based on electron spin ensembles (for example comprising one of the spin systems described in \Sec{sec:systems}), which are able to store and recall the states of single microwave photons. Such memories benefit from the long coherence lifetimes found in spin ensembles, and are intended to interface with quantum processors operating in the microwave domain (e.g.\ those based on superconducting qubits~\cite{Kubo2011} or other types of spin qubits). Crucially, through the use of inhomogeneous broadening (either natural, or created through magnetic field gradients), many separate modes of spin excitation can be addressed, enabling an ensemble of $N$ spins to store $\mathcal{O}(\sqrt{N})$ qubits~\cite{Wesenberg2009,Wu2010,Julsgaard2013,Grezes2014}.

The essence of the microwave quantum memory protocol is summarised in \Fig{fig:qmemory}, and comprises a spin ensemble which is strongly coupled to a microwave resonator. To enter this regime, the coupling between the spin ensemble and the microwave cavity, $g_{\rm ens}$,  must exceed both the spin linewidth ($\gamma_2^*$) and the coupled cavity linewidth ($\kappa = \omega_c/Q$), where $Q$ and $\omega_c$ are the cavity the Q-factor and frequency, respectively. The ensemble coupling $g_{\rm ens}$ derives from the coupling strength of a single spin to the cavity $g_0$, enhanced by a factor $\sqrt{N}$ for $N$ identical spins (in practice, the spin-coupling strength can vary across the ensemble and $g_{\rm ens}$ is obtained from a numerical integration). The single spin-cavity coupling $g_0$ 
%
can 
be determined by first calculating the magnitude of the zero-point fluctuations of the current in the resonator (i.e.\ the current corresponding to a resonator energy of $\frac12\hbar\omega_c$), then calculating the magnetic field produced by such a current, and the coupling strength of the relevant spin transition to this field. 

To reach the `strong coupling' regime ($g_{\rm ens} \gg \kappa,\gamma_2^*)$, a large cavity Q-factor is therefore desirable and this is typically achieved using structures made from superconducting materials such as Al, Nb, NbN, NbTiN and TiN.
However, superconductivity is only sustained up to a certain value of DC magnetic field, $B_c$ (which varies according to the material), and Q-factor degradation of a superconducting resonator is observed at fields well below $B_c$, imposing restrictions on the field which can be applied to the spins to obtain a suitable ESR transition frequency.
In addition, the spin concentration must be optimised to ensure a spin ensemble with a narrow linewidth (and long coherence time), while obtaining a significant $\sqrt{N}$-enhancement of the spin-cavity coupling. For this reason, the use of spins tuned to clock transitions as the storage medium for the microwave quantum memory is particularly attractive as it enables the spin concentration to be increased while reducing the impact on spin-decoherence caused by spin-spin interactions.

\subsection{Resonator designs}
\label{sec:resonators}
\begin{figure}
  \centering
  \includegraphics[width=90mm]{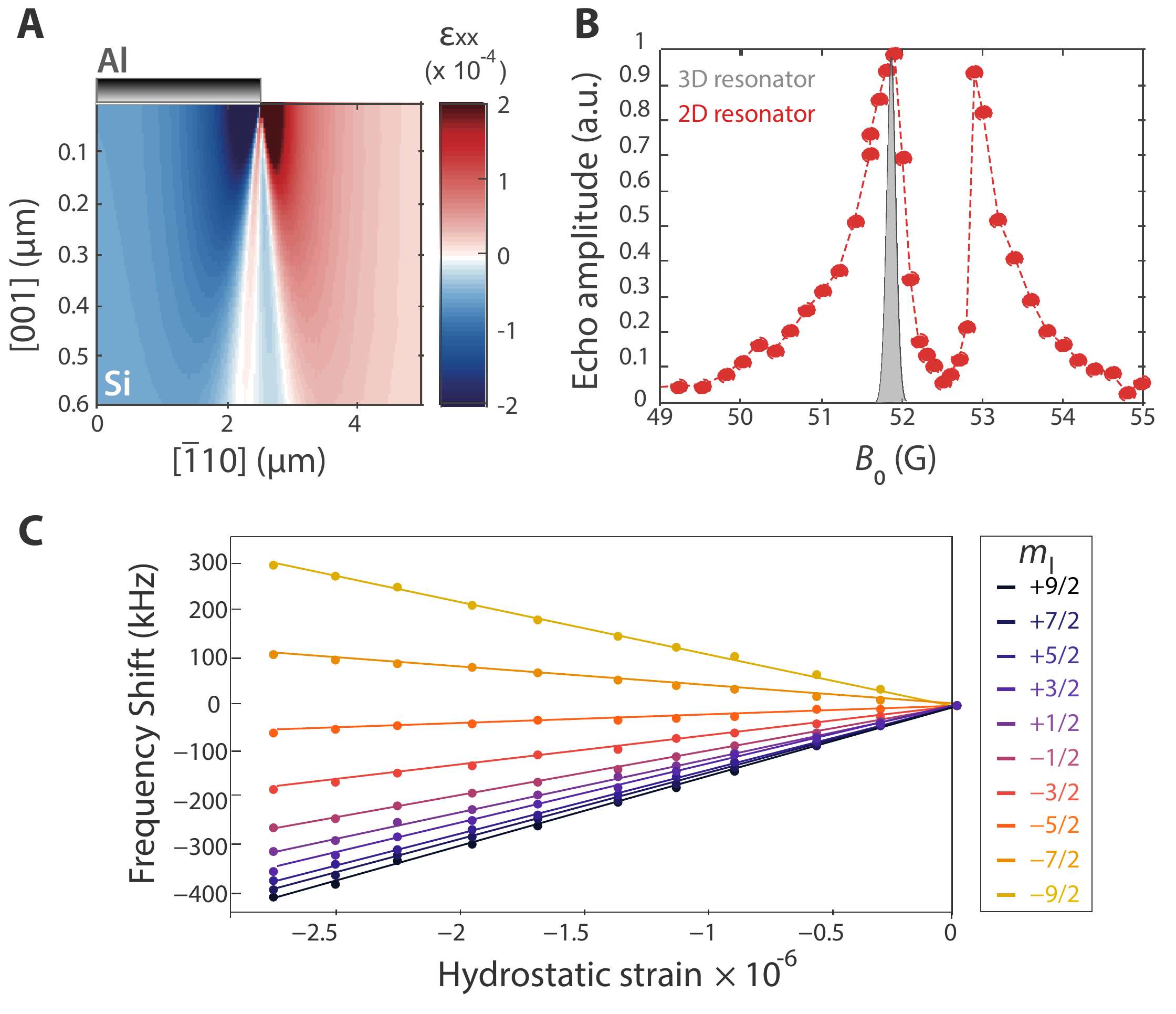}
  \caption{(A) Finite element modelling of strain induced within silicon underneath the edge of a patterned aluminium film, at low temperatures, caused by the mismatch of thermal expansion coefficient between the two materials. Al film thickness is taken as 50~nm, and the strain plotted is the component along the [100] silicon crystal orientation. (B) The resulting strain can cause major changes to the ESR lineshape of spins within the silicon. This panel shows the ESR lineshapes for Bi donors in \sitwoeight, measured using the pristine wafer (with no aluminium) in a 3D ESR resonator, and using the 2D resonator formed of a patterned Al film deposited on the silicon wafer (see \Fig{fig:resonators}D). The strain within the silicon varies from compressive (underneath the resonator) to tensile (beyond it), leading to a substantially broadened and split lineshape, caused by a modification of the Bi:Si hyperfine coupling strength. (C) ESR frequency shifts of Bi donors in Si have been studied in separate experiments applying a defined strain to the silicon, highlighting that relatively small strains ($\mathcal{O}(10^{-6})$) are sufficient to lead to ESR line broadening. 
  Panels (A) and (B) are adapted from Ref~\cite{Pla2017}, and (C) is adapted from Ref~\cite{Mansir2017}.
  }
  \label{fig:strain}
\end{figure}

For 3D cavities at X-band, typical values of single spin-cavity coupling, $g_0$, are $2\pi\times50$~mHz with a cavity mode volume of $\mathcal{O}(0.1)$~cm$^3$~\cite{Abe2011}. Given a spin concentration of $10^{15}$~cm$^{-3}$, and assuming full spin polarisation, this would lead to a total $g_{\rm ens}/2\pi \sim0.5$~MHz, requiring $Q>20,000$ to achieve strong coupling. Similar values of $g_{\rm ens}$ are achievable with planar superconducting resonators based on a $\lambda/4$ or $\lambda/2$ transmission line~\cite{Schuster2010, Sigillito2017}, or a lumped-element design where a capacitor and inductor are formed from a patterned superconducting film~\cite{Bienfait2016} (see \Fig{fig:resonators}). An advantage of the planar approach is that Q-factors in the range of $10^5$ to $10^6$ can be readily achieved, even in the presence of a magnetic field, by aligning the field in the plane of the superconducting film~\cite{Samkharadze2016}. However, such advantages should be balanced against potential ESR line broadening of the sub-surface spins caused by the planar resonator. For example, the ESR linewidth of bismuth donor spins in silicon was significantly increased due to mechanical strain created in the silicon by the aluminium lumped element resonator, due to differences in thermal expansion coefficient between the materials (see \Fig{fig:strain})~\cite{Bienfait2016}.

\begin{figure*}
  \centering
  \includegraphics[width=120mm]{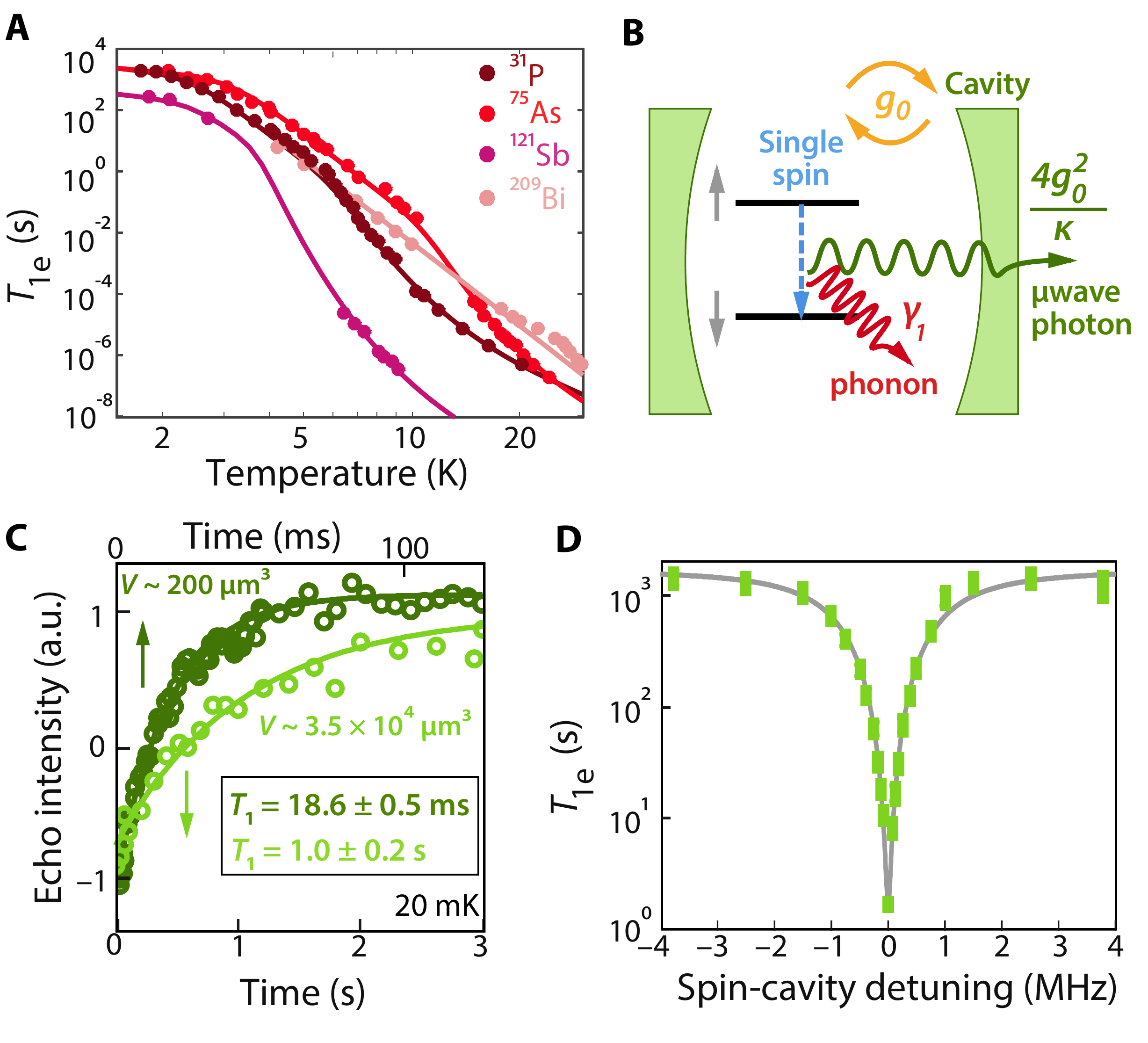}
  \caption{(A) The electron spin relaxation time, \tone\ of Group V donor spins in silicon exceed one second at temperatures below 5--10~K, and rise to approach or exceed hours at temperatures below 2~K. Many other electron spin systems also exhibit long \tone\ at sub-K temperatures. For this reason, it is important to identify methods to rapidly relax such spins to thermal equilibrium if they are to be coupled to superconducting cavities operating at typtical temperatures of tens of milliKelvin. (B) By placing the spin in a cavity, the rate of spin relaxation by spontaneous emission of a microwave photon can be substantially enhanced to a rate equal to $4g_0^2/\kappa$. If this exceeds the intrinsic (e.g. spin-lattice) relaxation rate $\gamma_1$, then the cavity-induced spin relaxation becomes the dominant mechanism, and returns the spin polarisation to thermal equilibrium. (C) This is demonstrated for the case of Bi donors in Si, coupled to a high-Q superconducting resonator (see \Fig{fig:resonators}D). Two inversion-recovery experiments are shown, obtained from the same sample measured using resonators with different mode volumes, with a corresponding difference in cavity-induced spin relaxation time observed (the loaded Q-factors were $3\cdot10^5$ and $4\cdot10^4$ for the larger and smaller resonators, respectively. (D) By detuning the spins from the cavity frequency, the degree of cavity-induced spin relaxation can be engineering, yielding a variation of three orders of magnitude in \tone, with just 2~MHz of detuning. 
  Panel (A) adapted from Ref~\cite{GaryThesis} and panels (C,D) adapted from Refs~\cite{Bienfait2016} and~\cite{Probst2017} .
  }
  \label{fig:purcell}
\end{figure*}

A set of CPW resonators can be capacitively coupled to a feedline (see \Fig{fig:resonators}A~\cite{Schuster2010}), allowing several different frequencies to be probed simultaneously. Alternatively, a $\lambda/2$ resonator may be formed by terminating both ends with Bragg mirrors which are both coupled to microwave ports (see \Fig{fig:resonators}B~\cite{Sigillito2017}). The Bragg mirror is formed by modulating the impedance of the CPW (through its width), creating a photonic bandgap in the microwave domain around the frequency of the resonator. The key advantage of this approach is that it enables other frequencies to be readily transmitted through the CPW, allowing ENDOR~\cite{Sigillito2017} or DEER~\cite{Asfaw2017} experiments to be performed. A second advantage is that, due to its ability to transmit DC, such a photonic bandgap resonator can be frequency-tuned by applying a current (see \Sec{sec:tuning}). 

Lumped-element superconducting microwave resonators can be created by defining inductive and capacitive elements on the superconducting film, often using an interdigitated structure or gap for the capacitor, and a wire inductor ranging from a few microns in width down to a few hundred nanometres. Figures~\ref{fig:resonators}D and \ref{fig:resonators}E illustrate two contrasting resonator designs, respectively optimising for low-impedance and high-impedance structures. For inductive coupling to spins, a low-impedance design ensures the vacuum fluctuations of the resonator yield larger currents, which increases the spin-cavity coupling $g_0$~\cite{Eichler2017}. Conversely, for electrically driven spin resonance (for example, on systems with significant spin-orbit coupling or double quantum dots with strong field gradients), a \emph{high}-impedance resonator is preferable, as this maximises the magnitude of the zero point \emph{voltage} fluctuations in the resonator~\cite{Samkharadze2016}.

\subsection{Purcell effect}
A microwave quantum memory should be maintained at sufficiently low temperature to avoid thermal excitation of the electromagnetic field in the cavity, which requires $T \ll \hbar\omega_c/k_B$ (e.g.~$T\ll0.5$~K for a frequency of 10~GHz). At such low temperatures, the spin relaxation time can become rather long (e.g.~hours for donor electron spins in silicon --- see \Fig{fig:purcell}A) and so it is beneficial to have the ability to `reset' the spins, accelerating their return to thermal equilibrium. Some techniques exist to achieve this optically, exploiting specific optical transitions within the spin defect or host material. However, a general method for enhancing spin relaxation times is based on the Purcell effect~\cite{Purcell1946}, in which spontaneous emission of a two-level system placed in a resonant cavity is accelerated by a factor $3Q\lambda^3/4\pi^2V$ (where $V$ is the mode volume of the cavity and $\lambda$ the radiation wavelength).

Cavity-enhanced spontaneous emission (or `Purcell relaxation') has been used extensively with optical and microwave cavities coupled to electric dipole transitions~\cite{Goy1983}, however, its application in enhancing electron spin relaxation was only recently achieved~\cite{Bienfait2016}. The inherently low spontaneous emission rates for electron spins (about 10,000 years for a free electron spin in a 0.3~T field) mean that Purcell enhancement factors of $\mathcal{O}(10^{11})$ or greater are required to enable cavity-induced spin relaxation to compete with the other spin relaxation mechanisms based on (e.g.) spin-phonon coupling. Such factors can be achieved using microwave resonators with $Q\sim10^5$ and $V\lesssim 10^6~\upmu$m$^3$, assuming a frequency of 10~GHz~\cite{BienfaitThesis2017}. The overall cavity-enhanced relaxation rate can be conveniently expressed as $\Gamma_{\rm P}=4g_0^2/\kappa$ (see \Fig{fig:purcell}B) --- when this exceeds $\gamma_1$, the intrinsic spin relaxation rate, we enter the `Purcell regime' where the spin relaxation is determined by the cavity (see Table~\ref{table:coupling}). As seen in \Fig{fig:purcell}D, the Purcell regime allows the spin relaxation rate to be controlled in-situ by changing the spin-cavity detuning, providing a versatile `reset switch' for spins at low temperatures, as well as providing potential applications in dynamical nuclear polarisation where the relaxation rates of a particular transition can be selectively enhanced~\cite{Bienfait2016}. Recent results with smaller cavities ($V\sim200~\upmu$m$^3$) have demonstrated single-spin coupling strengths of around 400~Hz, leading to an observed cavity-induced spin relaxation time of about 20~ms for Bi donors in Si, despite the lattice temperature being 20~mK~\cite{Probst2017}.

\begin{figure*}
  \centering
   \includegraphics[width=170mm]{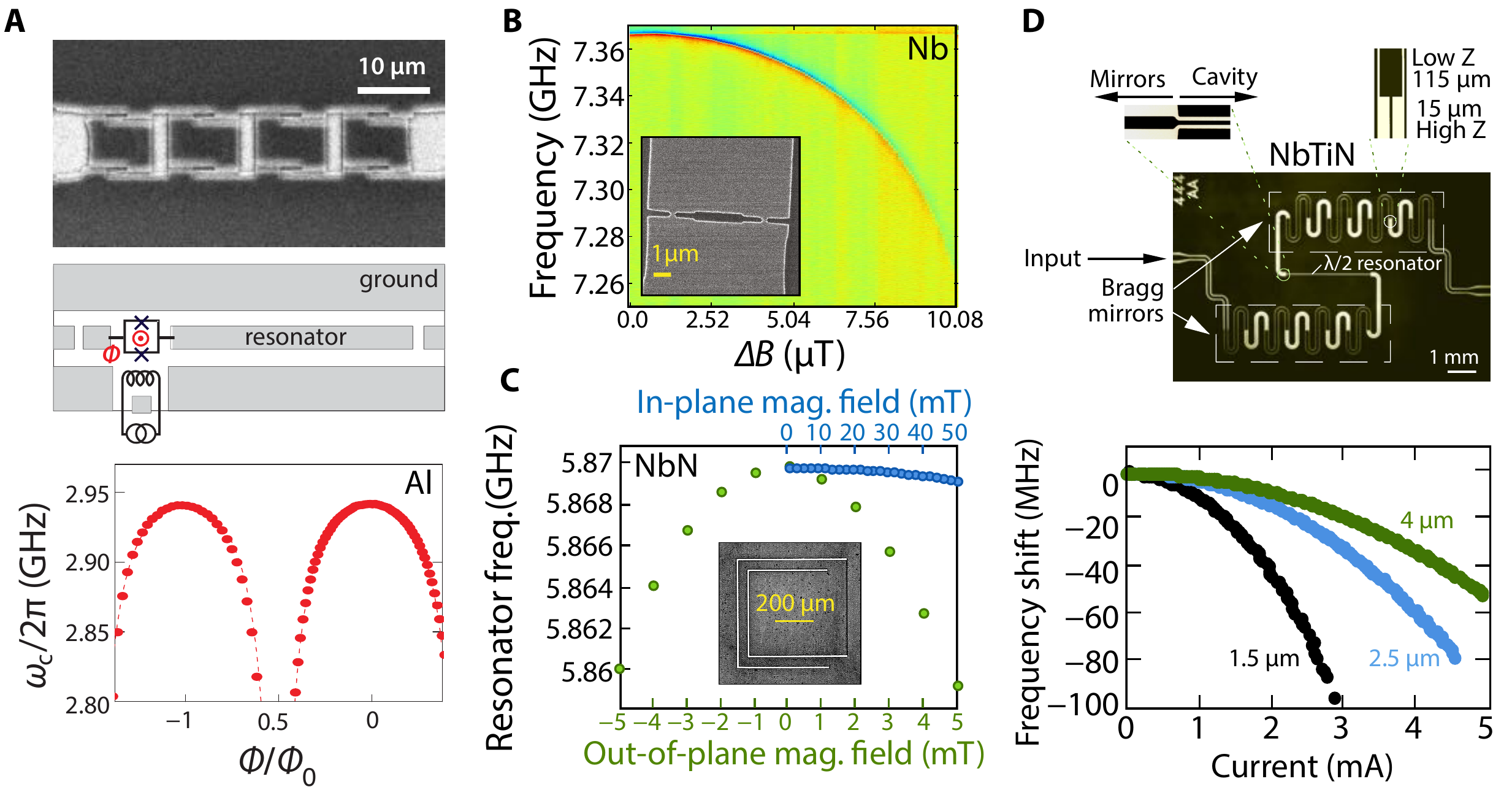}
  \caption{Controlling the spin-cavity detuning is important for several reasons, including gating the spin-cavity coupling in quantum memory applications, engineering spin-relaxation, and performing high-sensitivity ESR. Various methods to achieve this exist, depending on the type of materials used for the superconducting resonator. (A) Top panel is an optical micrograph displaying a series of SQUID loops, made using Al/Al$_2$O$_3$/Al Josephson junctions, which act as a magnetic-field-tunable inductor that can be incorporated within a lumped element or CPW resonator. The SQUID loops can be tuned by a globally applied bias field, or by a local bias loop on-chip, as shown in schematic (middle panel). In this case, 150~MHz tuning of the resonator frequency can be achieved with a bias current of about 1~$\upmu$A (bottom panel).. 
  (B) In other materials, such as Nb, SQUIDs can be formed using narrow constrictions of superconductor, in this case (20~nm). When incorporated into a CPW resonator such a SQUID enables tuning of the resonator frequency by 100~MHz with 10~$\upmu$T of applied out-of-plane field.
  (C) The use of SQUIDs can be avoided, instead relying on the change in kinetic inductance in the superconducting film caused by an applied out-of-plane magnetic field.This is demonstrated for the case of a NbN lumped element resonator, whose frequency is tuned by about 10~MHz using an 5~mT out-of-plane field, while maintaining $Q>10^5$. 
  Panel (A) is adapted with permission from Ref.~\cite{Kubo2010}, \textcopyright~American Physical Society; 
  (B) from Ref~\cite{Kennedy2017}, and (D) from Ref~\cite{Asfaw2017} with the permission of AIP Publishing.
  }
  \label{fig:tuning}
\end{figure*}

\renewcommand{\arraystretch}{2}
\begin{table}
\begin{center}
\begin{tabular}{|p{3cm}|p{5cm}|}
\hline
$g_{\rm ens} > \{\kappa,\gamma_2^*\}$ & Strong coupling (ensemble)\\  
\hline
$\frac{4g_{\rm ens}^2}{\kappa} > \gamma_2^*$ & High cooperativity (ensemble) \\  
\hline
$\frac{4g_0^2}{\kappa} > \gamma_2$ & High cooperativity (single spin) \\  
\hline
$\frac{4g_0^2}{\kappa} > \gamma_1$ & Purcell regime \\  
\hline
\end{tabular}
\caption{Various regimes of interest for one or more spins coupled to a cavity.}
\label{table:coupling}
\end{center}
\end{table}

It is worth pausing to consider the prospects for extending these results to cavity-enhanced relaxation of nuclear spins, indeed this is the goal which motivated Purcell's original proposal~\cite{Purcell1946}. Intrinsic nuclear spin relaxation times of hours or longer are common at low temperatures, such that a $\Gamma_{\rm P}$ of $10^{-4}$~Hz would already be sufficient to dominate spin relaxation in certain systems. $\Gamma_{\rm P}$ scales with the gyromagnetic ratio squared, such that if a cavity similar to that shown by Probst~\emph{et al.}~\cite{Probst2017} with $V\sim200~\upmu$m$^3$ were coupled to $^1$H spins, one could expect $\Gamma_{\rm P}\sim 10^{-4}$~Hz. Microwave resonators have been shown to maintain $Q>10^5$ under magnetic fields up to 6~T~\cite{Samkharadze2016}, corresponding to a \oneh~NMR frequency of about 250~MHz, however, there would be challenges in designing a resonator at such a low frequency while maintaining a similarly high Q-factor and small mode volume. Finally, there is the challenge of detecting the signal from the number of nuclear spins within a volume as small as a few hundred $\upmu$m$^3$.

\subsection{Resonator tuning}
\label{sec:tuning}
Bringing the cavity onto resonance with the relevant ESR transitions of the spin system is a prerequisite to benefit from cavity-enhanced spin relaxation or indeed perform high-sensitivity ESR as discussed further in \Sec{sec:esr}. An obvious way to control the spin-cavity detuning is to apply a magnetic field, using the Zeeman interaction to bring the spins onto resonance with the cavity. This can be used, for example, to compensate for the difference in the actual frequency of the fabricated resonator and that it was designed for. The applied magnetic field also shifts the resonator frequency, but if the field is kept in the plane of the superconducting film, this effect is negligible compared to the typical electron gyromagnetic ratio of 28~GHz/T. 

To achieve microwave photon storage in a spin-based quantum memory, it is necessary to obtain spin-cavity coupling for a precisely determined period of time, requiring the ability to quickly tune the spins and cavity into and out of resonance~\cite{Kubo2011}. In such cases, a globally applied magnetic field would need to be applied through a secondary set of coils with lower inductance to allow faster switching. Faster, more sensitive tuning of the resonator frequency can be achieved using one or more SQUIDs, incorporated into the resonator structure (see~\Fig{fig:tuning}A,B), whose inductance is sensitive to the perpendicular magnetic flux density. Resonator frequency shifts of order 100~MHz can be achieved using perpendicular fields applied globally (e.g.~of 10~$\upmu$T)~\cite{Kubo2010}, or created by on-chip bias loops (e.g.~with 1~$\upmu$A of current)~\cite{Kennedy2017}.  

A final method to achieve fast control of resonator frequency makes use of photonic bandgap NbTiN resonators based on Bragg mirrors, where DC currents can be directly passed through the resonator to shift its frequency. This has been demonstrated using resonators with $Q\sim3000$, where a DC current of 3~mA resulted in a frequency shift of about 100~MHz, due to the impact of the bias current on the kinetic inductance~\cite{Asfaw2017}.

\section{High-sensitivity pulsed ESR with superconducting devices}
\label{sec:esr}
Much of the development around microwave quantum memories, described above, can also be directed towards the improvement of spin number sensitivity in ESR. Various derivations for the sensitivity of pulsed ESR have been performed~\cite{Prisner1994,Rinard2002}, though a simple and useful expression for the minimum number of spins detectable in a single echo can be obtained by considering simple arguments from microwave quantum optics, assuming the cavity damping rate is dominated by external coupling and internal cavity losses are negligible (see Ref~\cite{Bienfait2015} for a more detailed derivation).

We consider an ensemble of $N$ spins each with identical coupling to the cavity ($g_0$), at the end of a pulsed ESR experiments where a spin echo is formed. At this point, the spins are in-phase and hence there is an enhanced collective coupling of the ensemble to the cavity of $g_{\rm ens}=\sqrt{N}g_0$. Each spin now experiences an enhanced cavity-induced relaxation rate of $4g^2_{\rm ens}/\kappa$, leading to microwave photon emission from the cavity over some duration approximately equal to the spin ensemble dephasing time (1/$\gamma_2^*$). Thus, the total number of microwave photons emitted by the spins during a spin echo is:
\beq
N_{\rm photons}\sim N\frac{4g^2_{\rm ens}}{\kappa\gamma_2^*}=N^2\frac{4g^2_0}{\kappa\gamma_2^*}=N^2C_0,
\eeq
where $C_0$ is the single spin cooperativity. The amplitude of the microwave signal can be obtained from the square root of the photon number, and we can account for finite spin polarisation, $p$, by taking the effective number of fully-polarised spins to be $pN$. Therefore the signal-to-noise ratio (SNR) from a single echo can be written as:
\beq
{\rm SNR_{\rm echo}} \sim \frac{pN\sqrt C_0}{\sqrt{n}},
\eeq
where $n$ is the number of noise photons added to the signal. The minimum detectable number of spins (i.e.~that for which the SNR is equal to one) is then:
\beq
N_{\rm min} \sim \frac{\sqrt n}{p\sqrt C_0},
\eeq
which is minimised by reducing the number of noise photons and by increasing the spin polarisation,  cavity Q-factor,  spin dephasing time, and single spin-cavity coupling.

\begin{figure*}
  \centering
   \includegraphics[width=170mm]{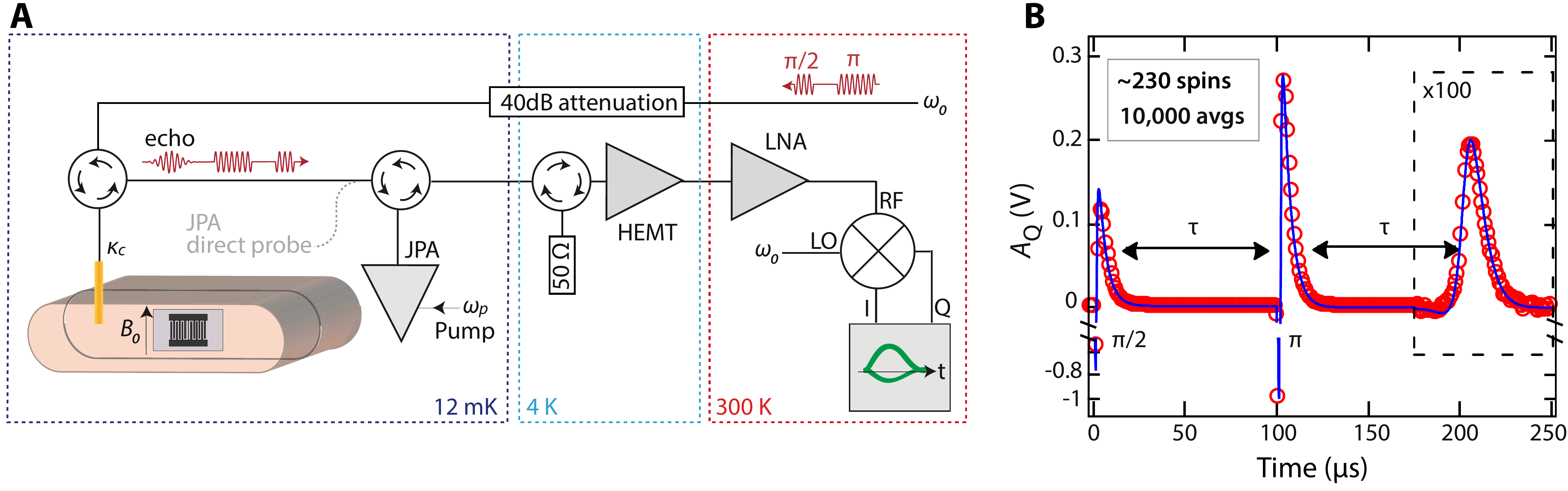}
  \caption{(A) Schematic showing experimental set-up for high-sensitivity ESR at milliKelvin temperatures. Microwave pulses are sent down an attenuated line to the base temperature of a dilution fridge, and the reflected pulses and emitted spin echo are passed to a quantum-limited Josephson Parametric Amplifier (JPA), which operates in reflection. The amplified signal is then further amplified at 4~K using a high electron mobility (HEMT) amplifier, and again at 300~K using a low noise amplifier (LNA), before being analysed through an IQ-mixer. (B) Example two-pulse Hahn echo experiment measured on such a set-up, using a microresonator where $\sim230$~spins are on-resonance, contributing to the echo signal. The signal-to-noise is 0.9 per single-shot echo. 
 Figure adapted from Refs~\cite{Bienfait2015} and \cite{Probst2017}.
  }
  \label{fig:ESRsensitivity}
\end{figure*}

The contribution to $n$ from thermal noise goes as $\frac12\coth{\frac{\hbar\omega_0}{2k_BT}}$, which tends to 1/2 at low temperatures. Recent developments motivated by measuring superconducting qubits have led to significant improvements in so-called Josephson Parametric Amplifiers (JPAs), which are loss-less, quantum-limited amplifiers, operating at microwave frequencies (e.g.\ 1-12 GHz)~\cite{Yaakobi2013,Zhou2014}. They can be incorporated into an ESR detection arm, mounted at milliKelvin temperatures (see \Fig{fig:ESRsensitivity}) and produce no further noise photons beyond thermal noise (when operated in the mode where only one signal quadrature is amplified). Therefore, for ESR at milliKelvin temperatures where spins are highly polarised, thermal noise is determined by vacuum fluctuations, and loss-less amplifiers can be used, $N_{\rm min}$ can be approximately written as $1/\sqrt{2C_0}$.

The superconducting resonators described in \Sec{sec:resonators} have shown Q-factors above $10^5$ and demonstrated spin-cavity coupling $g_0/2\pi=450~$Hz~\cite{Probst2017}. When used to perform ESR of Bi donors in Si with \ttwostar~$\sim5~\upmu$s, a sensitivity of $N_{\rm min}=260$~spins was demonstrated, in good agreement with the expected value. Furthermore, the use of the Purcell effect to determine spin relaxation means that a repetition rate of at least 16~Hz can be used regardless of the intrinsic relaxation times of the spin system, such that a sensitivity of 65~spins/$\sqrt{\rm Hz}$ may be inferred (see \Fig{fig:ESRsensitivity}B). The primary factor in improving the sensitivity further is likely to come from shrinking the resonator volume further, increasing $g_0$. In addition to improving the SNR per echo, this also enables faster repetition rates due to the Purcell effect, such that the ESR sensitivity per $\sqrt{\rm Hz}$ follows $g_0^{-2}$~\cite{Haikka2017}.

Despite these impressive opportunities in pulsed ESR sensitivity, there are several practical issues arising from the use of such resonators and milliKelvin environments which are far from those typically used in conventional ESR. First, the very large Q-factors of $10^5$ or more may not be well-suited for samples with short coherence times (requiring short pulses). In such cases, the resonators could be over-coupled, or other resonators used, bearing in mind a 1000-fold reduction in the Q-factor would reduce the per-echo sensitivity by a factor of about 30, but the per-root-Hz sensitivity by 1000. The possibility of Q-switched resonators, or shaped microwave pulses to obtain short drive pulses in high-Q cavities~\cite{Franck2015}, could have useful applications in such situations. 

Second, when performing magnetic field sweeps, as is typical in ESR, it is important to consider the shift in the resonator frequency (and, potentially Q-factor) which may remain even if the field is well aligned with the plane of the resonator (see \Sec{sec:tuning}). Such effects could be accounted for, to some extent, in the post-processing of the data. Alternatively, it may be possible to compensate for field-induced shifts of the resonator frequency through some other tuning mechanism, for example, using a separately biasable SQUID, or by passing a DC current through a photonic bandgap resonator. 

Third, it is worth considering the microwave powers used in such experiments. In order to ensure the thermal noise at the sample is not dominated by black-body radiation from room temperature, the input line feeding the microwave pulses must be heavily attenuated at various stages in the cryostat. However, this should not raise particular concern because the small resonator volume enables extremely low powers to be used. For example, in Ref~\cite{Probst2017} a $\pi$-pulse of duration 1~$\upmu$s was achieved using 0.5~pW input power into the resonator, while $\pi$-pulse durations of 10~ns (more typical in ESR) would require 5~nW. 

A final practical consideration is the loading of the sample into the resonator volume: to take full advantage of the small values of $N_{\rm min}$ demonstrated, it is necessary to ensure there is an efficient sample preparation procedure with minimum `wasted' spins outside the resonator volume. In principle, this could be achieved using a microfluidic device to enable a small volume of injected material to reside over the resonator structure. 

\section{Summary}
A rich set of electron spin systems are available for use in the storage of quantum information, each with different advantages. Engineered quantum dots in semiconductor devices offer strong possibilities for large-scale integration for quantum information processing, a variety of optically-active defects and ions in solid-state materials offer opportunities to interface with light for measurement and microwave-optical conversion, while some impurities in suitable host environments offer electron spin coherence times exceeding seconds. Through the toolbox of cavity quantum electrodynamics, such long-lived spins can be coupled to high-Q superconducting resonators to build multi-mode memories for microwave photons. Preliminary results have already shown the exchange of a single qubit state between a spin ensemble and superconducting qubit~\cite{Kubo2011}, as well as multi-mode storage using weak microwave pulses~\cite{Wu2010,Grezes2014}, suggesting that a high-fidelity spin-based quantum memory could be demonstrated in near future. 
Finally, we have seen how parallel developments based on a similar experimental and theoretical toolbox have yielded improvements in pulsed ESR sensitivity, with the current state-of-the-art at 65~spins/$\sqrt{\rm Hz}$. Given the advantages arising from the high spin polarisation, low thermal noise, high-Q superconducting resonators and quantum-limited amplifiers, there are good reasons to expect milliKelvin temperatures will become increasingly common in ESR laboratories.
	
\section{Acknowledgements}
We acknowledge the support of the European Research Council under the European CommunityÕs Seventh Framework Programme (FP7) through Grant Agreements No. 615767 (CIRQUSS) and No. 279781 (ASCENT), the European UnionÕs Horizon 2020 programme under Grant Agreement No. 688539 and of the Agence Nationale de la Recherche through the project QIPSE.

\bibliography{library}
		
\end{document}